\documentclass[journal = apchd5]{achemso} \setkeys{acs}{articletitle = true}
\usepackage{graphicx}
\usepackage{color}
\usepackage{amsmath}
\usepackage{braket}
\usepackage{gensymb}
\usepackage[normalem]{ulem}
\usepackage{lineno}

\title{Electric field correlation measurements on the electromagnetic vacuum state} 

\author{Ileana-Cristina  Benea-Chelmus}%
\email{ileanab@ethz.ch}
\affiliation{ETH Zurich, Institute of Quantum Electronics, Auguste-Piccard-Hof 1, Zurich 8093, Switzerland}

\author{Francesca Fabiana Settembrini}
\affiliation{ETH Zurich, Institute of Quantum Electronics, Auguste-Piccard-Hof 1, Zurich 8093, Switzerland}
\author{Giacomo Scalari}
\affiliation{ETH Zurich, Institute of Quantum Electronics, Auguste-Piccard-Hof 1, Zurich 8093, Switzerland}
\author{J\'er\^ome Faist}
\email{jerome.faist@phys.ethz.ch}
\affiliation{ETH Zurich, Institute of Quantum Electronics, Auguste-Piccard-Hof 1, Zurich 8093, Switzerland}
\date{\today}
\RequirePackage[normalem]{ulem} 
\RequirePackage{color}\definecolor{RED}{rgb}{1,0,0}\definecolor{BLUE}{rgb}{0,0,1} 
\providecommand{\DIFaddbegin}{} 
\providecommand{\DIFaddend}{} 
\providecommand{\DIFdelbegin}{} 
\providecommand{\DIFdelend}{} 
\providecommand{\DIFaddbeginFL}{} 
\providecommand{\DIFaddendFL}{} 
\providecommand{\DIFdelbeginFL}{} 
\providecommand{\DIFdelendFL}{} 
\newcommand{\DIFscaledelfig}{0.5}
\RequirePackage{settobox} 
\RequirePackage{letltxmacro} 
\newsavebox{\DIFdelgraphicsbox} 
\newlength{\DIFdelgraphicswidth} 
\newlength{\DIFdelgraphicsheight} 
\LetLtxMacro{\DIFOincludegraphics}{\includegraphics} 
\newcommand{\DIFaddincludegraphics}[2][]{{\color{blue}\fbox{\DIFOincludegraphics[#1]{#2}}}} 
\newcommand{\DIFdelincludegraphics}[2][]{
\sbox{\DIFdelgraphicsbox}{\DIFOincludegraphics[#1]{#2}}
\settoboxwidth{\DIFdelgraphicswidth}{\DIFdelgraphicsbox} 
\settoboxtotalheight{\DIFdelgraphicsheight}{\DIFdelgraphicsbox} 
\scalebox{\DIFscaledelfig}{
\parbox[b]{\DIFdelgraphicswidth}{\usebox{\DIFdelgraphicsbox}\\[-\baselineskip] \rule{\DIFdelgraphicswidth}{0em}}\llap{\resizebox{\DIFdelgraphicswidth}{\DIFdelgraphicsheight}{
\setlength{\unitlength}{\DIFdelgraphicswidth}
\begin{picture}(1,1)
\thicklines\linethickness{2pt} 
{\color[rgb]{1,0,0}\put(0,0){\framebox(1,1){}}}
{\color[rgb]{1,0,0}\put(0,0){\line( 1,1){1}}}
{\color[rgb]{1,0,0}\put(0,1){\line(1,-1){1}}}
\end{picture}
}\hspace*{3pt}}} 
} 
\LetLtxMacro{\DIFOaddbegin}{\DIFaddbegin} 
\LetLtxMacro{\DIFOaddend}{\DIFaddend} 
\LetLtxMacro{\DIFOdelbegin}{\DIFdelbegin} 
\LetLtxMacro{\DIFOdelend}{\DIFdelend} 
\DeclareRobustCommand{\DIFaddbegin}{\DIFOaddbegin \let\includegraphics\DIFaddincludegraphics} 
\DeclareRobustCommand{\DIFaddend}{\DIFOaddend \let\includegraphics\DIFOincludegraphics} 
\DeclareRobustCommand{\DIFdelbegin}{\DIFOdelbegin \let\includegraphics\DIFdelincludegraphics} 
\DeclareRobustCommand{\DIFdelend}{\DIFOaddend \let\includegraphics\DIFOincludegraphics} 
\LetLtxMacro{\DIFOaddbeginFL}{\DIFaddbeginFL} 
\LetLtxMacro{\DIFOaddendFL}{\DIFaddendFL} 
\LetLtxMacro{\DIFOdelbeginFL}{\DIFdelbeginFL} 
\LetLtxMacro{\DIFOdelendFL}{\DIFdelendFL} 
\DeclareRobustCommand{\DIFaddbeginFL}{\DIFOaddbeginFL \let\includegraphics\DIFaddincludegraphics} 
\DeclareRobustCommand{\DIFaddendFL}{\DIFOaddendFL \let\includegraphics\DIFOincludegraphics} 
\DeclareRobustCommand{\DIFdelbeginFL}{\DIFOdelbeginFL \let\includegraphics\DIFdelincludegraphics} 
\DeclareRobustCommand{\DIFdelendFL}{\DIFOaddendFL \let\includegraphics\DIFOincludegraphics} 

\begin{document}

\newpage

\textbf{Quantum mechanics ascribes to the ground state of the electromagnetic radiation~\cite{Loudon:2000up} zero-point electric field fluctuations that permeate empty space at all frequencies. No energy can be extracted from the ground state of a system and, therefore, these fluctuations cannot be measured directly with an intensity detector. The experimental proof of their existence came thus from more indirect evidence, such as the Lamb shift~\cite{Lamb:1947,Bethe:1947,Fragner1357}, the Casimir force between close conductors~\cite{Casimir:1948dh,doi:10.1063/1.1665432,Wilson2011} or spontaneous emission~\cite{walls1994quantum,Loudon:2000up}. A direct method to determine the spectral characteristics of vacuum field fluctuations has been missing so far. In this work, we perform a direct measurement of the field correlation on these fluctuations in the terahertz frequency range  using electro-optic detection~\cite{Wu:1995ec} in a non-linear crystal placed in a cryogenic environment. We investigate their temporal and spatial coherence, which, at zero time delay and spatial distance, has a peak value of {$6.2\cdot 10^{-2}~V^2/m^2$}, corresponding to a fluctuating vacuum field~\cite{Riek:2015ju,Riek:2017gu} of {$0.25~V/m$}. With this measurement, we determine the spectral composition of the ground state of electromagnetic radiation which lies within the bandwidth of electro-optic detection.}

The spectral properties of the ground state of a quantum system intimately determine its behaviour. An optical cavity, for example, shapes the spectral density of states of the vacuum state and spontaneous emission is enhanced at its resonance frequency~\cite{Pur46}. Moreover, systems in which matter excitations are ultrastrongly coupled to light  \textcolor{black}{are characterised by new eigenstates called polaritons}~\cite{Scalari1323:2012,jkeller:2017,Bayer:2017ab} that are predicted to have a ground state populated with virtual photons~\cite{PhysRevB.72.115303,Guenter2009}. A method to measure the spectral properties of the electromagnetic ground state in-situ would provide a direct experimental test of this property predicted theoretically.

The first order correlation function of a classical field \textcolor{black}{$G^{(1)}(\tau, \delta \vec r) = \langle E^*(t,  \vec r) E(t+\tau,  \vec r+\delta \vec r) \rangle $}~\cite{Loudon:2000up}, where the brackets express the time and spatial averages, defines the temporal and spatial coherence properties of light and yields its power spectrum after a Fourier transformation. $G^{(1)}(\tau, \delta \vec r)$ is \textcolor{black}{thus} retrieved by measuring the mean intensity of the electric field interfering with a delayed and displaced version of itself.  \textcolor{black}{The time delay $\tau$  is typically given by different path lengths in the two arms of an interferometer.} 

\textcolor{black}{The description of an interferometer for quantum fields supposes the replacement of the classical electric field by normally ordered positive and negative frequency parts of the electric field operator. The first order correlation function of quantum fields yields $G^{(1)}(\tau, \delta \vec r) = \langle \hat E^{(-)}(t,  \vec r) \hat E^{(+)}(t+\tau,  \vec r+\delta \vec r) \rangle$~\cite{Loudon:2000up}, where $\langle \rangle$ represents the expectation value.} As such, the response of an interferometer on the vacuum state is clearly zero. We show in the following that the use of electro-optic detection~\cite{Wu:1995ec} to measure directly the correlation function of the complete vacuum electric field \textcolor{black}{$\hat E^{(-)}(t, \vec r)+\hat E^{(+)}(t, \vec r)$} at two space-time points \textcolor{black}{$(t, \vec r)$ and $(t+\tau, \vec r+\delta \vec r)$}, instead of an intensity measurement, allows the measurement of the field correlation function on the vacuum state. In electro-optic detection, an ultrashort near-infrared~(NIR) probe pulse enables the measurement of the instantaneous electric field of  free-running terahertz~(THz) electromagnetic waves by probing locally the birefringence induced in a detection crystal with second order non-linear susceptibility $\chi^{(2)}$. Using this technique, we have recently measured the first and second order correlation of classical fields with sub-cycle temporal resolution~\cite{BeneaChelmus:2016cl,BeneaChelmus:2017di}. 

In the present study, the detection crystal is permeated at any time by an infinity of (vacuum) electromagnetic modes of arbitrary wavevector $\vec k$ and corresponding frequency $\Omega$. The electro-optic measurement of the superposition of all these modes corresponds in the quantum picture to the measurement of an operator~\cite{Moskalenko:2015hw} that depends linearly on the multi-mode vacuum field, expressed as a sum of plane waves: 
\begin{equation} \hat S_{eo}(t, \vec r) = \sqrt{C}\sum_{\Omega}\sqrt{\frac{\hbar\Omega}{2\epsilon_0\epsilon_r V}}(\hat a(\Omega)R(\Omega) e^{-i(\Omega t-\vec k \vec r)}- h.c.) \end{equation}  

$\sqrt{C} =-i  r_{41}n^3 l\omega_p I_p /c_0$ where $c_0$ is the speed of light in vacuum, $\omega_p = 2\pi \times 375$~THz the frequency of the probe, $l$ the crystal length, n the refractive index at $\omega_p$, $r_{41}$ the electro-optic coefficient of the detection crystal. $\hat E_{THz,\Omega}(t, \vec r) = \hat E^{(-)}_{THz,\Omega}(t, \vec r)+\hat E^{(+)}_{THz,\Omega} (t, \vec r)= i\sqrt{\frac{\hbar\Omega}{2\epsilon_0\epsilon_r V}}(\hat a(\Omega) e^{-i(\Omega t-\vec k \vec r)}- h.c.)$ is the vacuum electric field operator of a THz mode at frequency $\Omega$ confined to the volume V of dielectric constant $\epsilon_r$. $R(\Omega)$ is introduced to describe the frequency dependent responsivity function  of electro-optic detection that accounts for the phase matching (sec.~1.5 of supplementary information). \textcolor{black}{$I_p$ is the (classical) mean intensity of the coherent probe with a corresponding instantaneous electric field $\hat E_p(t, \vec r) = E_p(t, \vec r) + \delta \hat E_p(t, \vec r)$. The probe contains its own multimode vacuum contribution $\delta \hat E_p(t, \vec r)$~\cite{Loudon:2000up,Moskalenko:2015hw}.}  \textcolor{black}{The latter introduces an (undesired) noise field $\delta \hat {\tilde E}(t, \vec r) = \frac{\delta\hat E_p(t, \vec r)}{E_p(t, \vec r)}$ into the measurement and $\hat S_{SN}(t, \vec r) = \sqrt C \delta \hat {\tilde E}(t, \vec r) $ is the noise equivalent electro-optic signal. The total measured electro-optic signal is thus $\hat S(t, \vec r)= \hat S_{eo}(t, \vec r)+\hat S_{SN}(t, \vec r)$\cite{Moskalenko:2015hw}}.

We define the electro-optic field correlation operator as the anti-commutator of the electro-optic operators at \textcolor{black}{$(t, \vec r)$ and $(t+\tau, \vec r+\delta \vec r)$: $ \hat G^{(1)}_{eo} (\tau, \delta \vec r) =-\frac{1}{2C}\{ \hat S_{eo}(t+\tau, \vec r + \delta \vec r),\hat S_{eo}(t, \vec r) \}$. The electro-optic field correlation operator is composed, besides normally ordered ladder operators, also of not normally ordered ladder operators~\cite{Loudon:2000up}. They are instrumental to the field correlation measurements performed on the vacuum field}.  Its expectation value $G^{(1)}_{eo} (\tau, \delta \vec r)$ at $\delta \vec r = 0 $ for a thermal state with a mean photon occupation number per mode $ \langle\hat  n(\Omega ) \rangle$ is the electro-optic field correlation function

 \begin{equation}
 \textcolor{black}{G^{(1)}_{eo} (\tau,0)= \sum_{\Omega}\frac{\hbar \Omega}{2\epsilon_0  \epsilon_r V}(1+ 2 \langle\hat  n(\Omega) \rangle) |R(\Omega)|^2\cos{\Omega\tau}.} 
\end{equation}
In the experiment, to this quantity is added the correlation between the noise equivalent electric fields at the probe frequency, \textcolor{black}{$\langle \delta \hat E(t, \vec r)\delta \hat E(t+\tau, \vec r) \rangle$. It converges to zero as the NIR vacuum fields are uncorrelated in the two detection modes and its variance is brought below the signal levels discussed here by measurement integration, as demonstrated by the Allan deviation plots~(sec.~3.2 of supplementary information).} 

For a vanishing THz photon population, the input state is the vacuum state $\ket{0}$, described by $ \langle\hat  n(\Omega) \rangle = 0$. The electro-optic correlation function yields in this case \textcolor{black}{$ G^{(1)}_{eo} (\tau,0) = \sum_{\Omega}\frac{\hbar \Omega}{2\epsilon_0  \epsilon_r V} |R(\Omega)|^2\cos{\Omega\tau}$. }

\begin{figure*}[hbtp]
  \centering 
 \includegraphics[width=12cm]{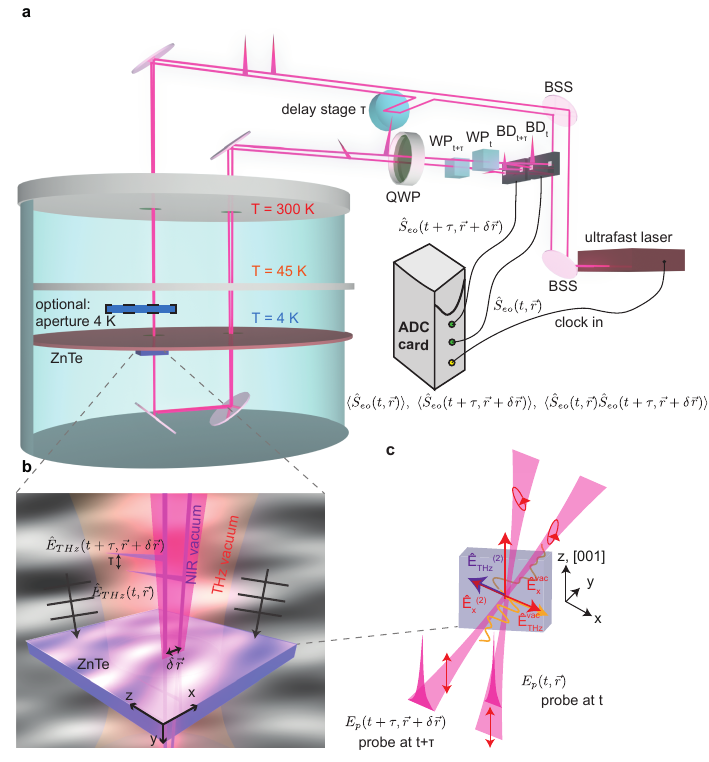}
  \caption{\textcolor{black}{\textbf{Experimental setup for the temporal and spatial electro-optic field correlation on vacuum and thermal fields} \textbf{a,}  Two probe pulses $E_{p}(t,\vec r)$ and $E_{p}(t+\tau,\vec r +\delta \vec r)$ sample the electric field of the propagating waves in the crystal with the repetition rate of the employed Mai Tai laser~(80~MHz). A mechanical delay stage provides a temporal delay of $\tau$ to one of the probe pulses. The detection crystal~(ZnTe, 110-cut) is placed inside a closed-cycle cryostat and is thermally anchored to the 4~K plate. It can be shielded from the blackbody radiation of the environment by two shields, with a temperature of 45~K and 4~K, respectively. The sampled electric fields introduce the electro-optic signals $S_{eo}(t,\vec r)$ and $S_{eo}(t+\tau,\vec r+\delta \vec r)$ on the two probe pulses, which are individually recorded by means of an analog-to-digital converter~(ADC). Here, the relevant correlations are computed in real-time. \textbf{b,}  The vacuum field fluctuations couple from the environment into the detection crystal, where the multimode electric field $ \hat E_{THz}(t,\vec r)$ and $ \hat E_{THz}(t+\tau,\vec r +\delta \vec r)$ is measured. An efficiently detected THz mode is composed of a superposition of plane waves which have favourable coherence properties with the probe beams. The lateral displacement of the pair of probe pulses $\delta x$ is determined by an external mirror.  \textbf{c,}  The vacuum fields polarised along the x axis of the crystal mix with the probes polarised along the z axis. These generated components along the x axis result in an elliptical polarisation of the probes after the crystal. QWP = quarter wave plate, WP = Wollaston prism, BD = balanced detector,  BSS = beam stabilisation system, ZnTe = zinc telluride.}}
  \label{fig:set}
\end{figure*}

\begin{figure*}[hbtp]
  \centering 
 \includegraphics[width=12cm]{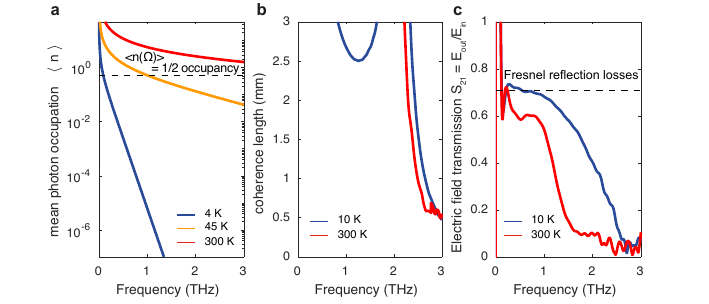}
  \caption{\textcolor{black}{\textbf{Mean photon occupation number per mode, coherence length and electric field transmission of THz radiation. }\textbf{a,}  Mean photon occupation number per mode for blackbody radiation at different temperatures~\cite{Loudon:2000up} . \textbf{b,} Coherence length of electro-optic detection for a detection crystal cooled to 10~K and 300~K. \textbf{c,} Electric field transmission through the 3~mm thick uncoated ZnTe detection crystal at different temperatures. }}
  \label{fig:set2}
\end{figure*}

In this work, we show experimentally that the electro-optic field correlation measurement on a vacuum state is non-zero. We implemented the two-point correlation measurement~\cite{BeneaChelmus:2016cl,BeneaChelmus:2017di} of $G_{eo}^{(1)}(\tau, \delta \vec r )$ using a pair of mode-matched 80~fs pulses of waist $w_0 = 125~\mu m$ that sample the multi-mode THz vacuum field in the two space-time points as shown in Fig.~\ref{fig:set}a and in the zoom~\ref{fig:set}b ($\delta \vec r$ is changed by steering mirrors outside the cryostat and $\tau$ is controlled via a mechanical delay stage). The probe polarisation \textcolor{black}{is oriented along the z axis of the zinc telluride~(ZnTe) detection crystal and thereby maximises the electro-optic effect but suppresses all undesired coherent $\chi^{(2)}$ effects~(sec.~1.1 and 1.6 of supplementary information). }The \textcolor{black}{3~mm thick} crystal is inserted in a cryostat that enables a control of its thermal environment with temperatures between 4~K and 300~K. \textcolor{black}{In this fashion, the contribution of each mode to the total electro-optic correlation function $G_{eo}^{(1)}(\tau, \delta \vec r )$ is controlled, by changing the mean population of the mode with thermal photons as shown in Fig.~\ref{fig:set2}a, the phase matching with the probe pulse, described by the coherence length shown in Fig.~\ref{fig:set2}b, and the absorption of THz photons in the detection crystal shown in Fig.~\ref{fig:set2}c.}  \textcolor{black}{The field induced birefringence is measured individually for each single probing pulse and separately for the two time-delayed trains of pulses with two balanced detection schemes, as shown in Fig.~\ref{fig:set}a. All measurements are saved and processed in real-time by means of a fast analog-to-digital converter that is phase locked to the laser oscillator at a repetition rate of $f_{rep} = 80~MHz$.}

\begin{figure*}[hbtp]
  \centering 
  \DIFaddendFL \includegraphics[width=12cm]{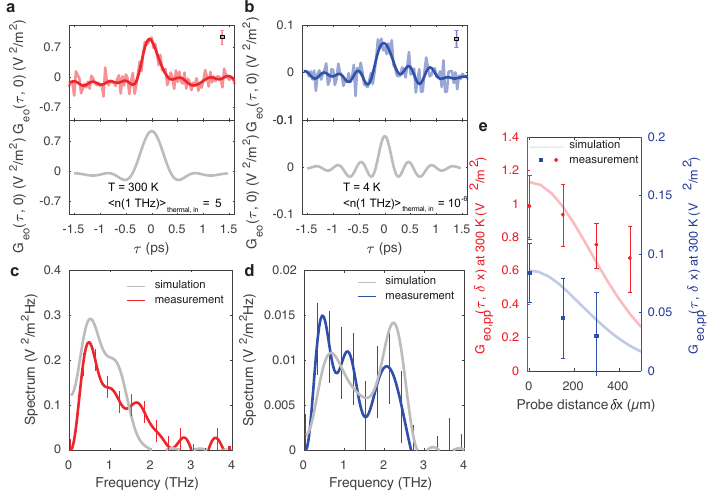}
  \caption{\textcolor{black}{\textbf{Electro-optic field correlation results at 300~K and 4~K.} }\textbf{a, - b,}  Electro-optic field correlation measurements $G_{eo}^{(1)}(\tau)$ (top plot) are compared to simulations (upper plot) for two temperatures, 300~K and 4~K, respectively. Faded lines denote raw measurements and thick lines the curves filtered by a low-pass Fourier filter of cut-off frequency of 3~THz, corresponding to the upper bound for efficient electro-optic detection in a 3~mm thick crystal. \textbf{c, - d,}  The power spectra of the detected fields are estimated by computing the real part of the Fourier transform of the raw (unfiltered) \textcolor{black}{electro-optic field correlations} $G_{eo}^{(1)}(\tau)$, owing to the fact that, by definition, the electro-optic correlation function is symmetric around $\tau = 0$. \textbf{e,}  Spatial coherence of the probed vacuum (blue squares) and thermal fields (red dots). The timetraces from which these values were extracted are shown in the supplementary information. \textcolor{black}{All error bars represent the  $1\sigma$ confidence interval.} }
  \label{fig:allG1}
\end{figure*}

The measured electro-optic field correlation $G_{eo}^{(1)}(\tau,0)$ is shown together with the associated power spectrum in Fig.~\ref{fig:allG1} for two distinct temperatures, 300~K and 4~K. They are compared to simulated results, computed using equation (2) as described in detail in sec.~1.3 and 1.4 of the supplementary information. 

In Fig.~\ref{fig:allG1}a~and~c we show results obtained when the system is at T~=~300~K. In this condition, the blackbody radiation from the environment dominates over the vacuum field, with a photon occupation number of for example $\langle\hat  n(\Omega = 1~THz) \rangle = 5$. The field coherence is preserved for a duration of 250~fs. The power spectrum reveals a large contribution of low frequency components, which exhibit a large mean occupation of thermal photons. The contribution of high frequency components is reduced by THz absorption in the detection crystal. The peak-peak signal  is {$G_{eo,pp}^{(1)}~=~0.98~V^2/m^2$}, and the root mean square of the noise is {$\sigma~=~0.134~V^2/m^2$}.

In Fig.~\ref{fig:allG1}b~and~d, we show results when the system is at 4~K base temperature~(optional 4~K aperture shown in Fig.~\ref{fig:set}a in use). In this condition, the trace and amplitude of the electro-optic field correlation changes dramatically and the thermal contribution is suppressed ($\langle\hat  n(\Omega = 1~THz) \rangle = 10^{-6}$). The power spectrum of vacuum fluctuations contains frequency components in the frequency band around 0.75~THz and around 2~THz. This \textcolor{black}{behaviour} is well reproduced \textcolor{black}{in our simulations}, and matches very well the coherence properties reported in Fig.~\ref{fig:set2}b. The peak-peak signal is {$G_{eo,pp}^{(1)}~=~0.084~V^2/m^2$}, and the root mean square of the noise is {$\sigma~=~0.018~V^2/m^2$}. To achieve this sensitivity, the integration time has been largely increased as discussed in sec.~3.2 and 3.3 of the supplementary information. 

We additionally investigate the spatial electro-optic field correlation of the probed vacuum and thermal fields by displacing the two probe beams in the crystal along the x axis, $\delta \vec r = \delta x \vec e_x$. \textcolor{black}{The lateral spatial coherence length~\cite{Loudon:2000up} of the probed multimode THz wave is determined by the participating waves with nonzero in-plane wavevector $\vec k_{\perp}$ and the transversal mode profile of the probe}. As shown in sec.~1.4 of the supplementary information, strictly speaking, in this case, the electromagnetic modes must be treated with their corresponding wavevector $\vec k$, rather than their frequency $\Omega$. In Fig.~\ref{fig:allG1}e, we report the peak-peak magnitude of $G_{eo}^{(1)}(\tau, \delta x )$ as a function of the probe spacing $\delta x$. We find that the spatial coherence is maintained over multiple wavelengths of the probed radiation, resulting in a lateral coherence length of $410~\mu$m at 300~K and \textcolor{black}{$375~\mu$m at 4~K}. We attribute the difference between our simulation and the experimental results to an uncertainty in the phase matching of the two waves. 

Finally, we demonstrate the pure electromagnetic origin of our measurements by removing the optional aperture shown in Fig.~\ref{fig:set}a when the system is cooled to 4~K. We change thus solely the properties of the detected light by allowing thermal radiation from the 45~K plate to reach the detection crystal.   \textcolor{black}{The electro-optic field correlation function, reported in Fig.~\ref{fig:45K}a, is considerably different from the one at 4~K. The peak-peak value is increased to $G_{eo,pp}^{(1)} = 0.14~V^2/m^2$. The coherence time exceeds several picoseconds.  In Fig.~\ref{fig:45K}b, we compare the detected photon number per mode extracted from the measured power spectrum to the expected photon number. We make use of equation (2) and the identical responsivity function $R(\Omega)$ in the two measurements (see methods). The good quantitative agreement proves hereby yet again the single photon level sensitivity of our measurements. }

Further insight into the quantum mechanical interpretation of our electro-optic correlation measurement on the vacuum state can be gained by comparing the latter to the conceptually equivalent case of two successive position measurements performed on a mechanical oscillator cooled to its ground state~\cite{Aspelmeyer:2014cea,PhysRevA.86.033840}. If the perturbation of the ground state due to the position measurement were to be neglected, these two measurements should remain uncorrelated because each one would give an independent Gaussian distribution of values representative of its wavefunction. Obviously, a single measurement of the electric field  perturbs however the vacuum state since the latter is not an eigenstate of the constituting ladder operators. In our opinion, the non-zero correlation we observe can be seen as arising from the perturbation of the measured vacuum state due to the first field measurement onto the second one, even if this perturbation is extremely small because we effectively perform a weak measurement. Notice that this is in complete agreement with the result of equation (2) in which the non-zero contribution to the correlation arises from the  not normally ordered ladder operators. To initiate this line of thoughts, we outline in sec.~1.2 of the supplementary material that electro-optic detection can be regarded as spontaneous parametric down conversion~\cite{PhysRevLett.75.4337} in a crystal with a $\chi^{(2)}$ non-linearity.  As shown in Fig.~\ref{fig:set}c, the initially uncorrelated x-polarised vacua at NIR and THz frequencies, $\hat E_{THz}^{vac}$ and $\hat E_{NIR}^{vac}$, are coherently amplified by the annihilation of a probe photon and the creation a new field $\hat E_{NIR}^{(2)} $, which introduces an elliptical polarisation of the initially linearly polarised probe as well as  $\hat E_{THz}^{(2)} $ which is generated and now contains real THz photons. However, future theoretical work is necessary to provide a comprehensive description of this picture.

In our measurements, we employ linear (field) detectors, which, in contrast to those used in superconducting circuits~\cite{daSilva:2010io,Bozyigit2010,Lahteenmaki2016} measure the electric field of a broadband wave with sub-cycle temporal resolution instead of the two quadratures of a narrowband one. In addition, our technique enables the measurement in a regime where the electro-optical quantum signals are much weaker than the single pulse shot noise, because the 1/f noise can be efficiently suppressed. We achieve a noise equivalent field squared of $1.8\cdot 10^{-2}~V^2/m^2$, which suffices to detect vacuum field fluctuations as well as fractions of thermal photons. If compared to the  \textcolor{black}{measurements} of Riek et al.~\cite{Riek:2015ju}, \textcolor{black}{our technique provides the spectral composition of the detected vacuum fields, as well as the lateral coherence length}. These characteristics may be essential to detect the emission of pure quantum light~\cite{2018arXiv180710519K} from non-adiabatic modulation of  a quantum system in the ultra-strong coupling regime~\cite{PhysRevB.72.115303}. While the present measurements, performed in an inorganic zinc telluride crystal, required long integration times to extract the signal, much larger signal-over-noise could be achieved using organics-based cavity enhanced THz detectors with extremely large electro-optic coefficients~\cite{Benea:2018}. By embedding the non-linear material in a resonator, the presence of a superradiant phase transition in a coupled light-matter system~\cite{Cong:16} as well as the influence of the vacuum field on the charge transport could be investigated~\cite{PhysRevLett.119.223601,Orgiu:213976}. In addition, by changing the detection of the near-infrared pulse from an ellipsometric to a projective polarisation measurement~\cite{Rungsawang:2008,vanKolck:2010}, electro-optic detection may provide a path for the generation of heralded single photons in the terahertz frequency range. 

\bibliographystyle{naturemag}
 \bibliography{bibtex-library}

\textbf{Acknowledgments} This work was funded by the European Research Council (Advanced Grant, Quantum Metamaterials in the Ultra Strong Coupling Regime) and the Swiss National Science Foundation (Grant 165639). We acknowledge the mechanical workshop at ETHZ. We acknowledge the contribution of Maryse Ernzer to the noise analysis tools, Dr. Elena Mavrona to the design of opto-mechanical components and the extraction of the refractive index of zinc telluride and Prof. Dr. Atac Imamoglu for fruitful discussions. We especially acknowledge both anonymous Reviewers for their extremely constructive questions and suggestions. 

\textbf{Author contribution} I.C.B.C. and J.F. conceived and designed the experiments. I.C.B.C., F.F.S and G.S. built the experimental setup. I.C.B.C. developed the data acquisition system and noise suppression protocols. I.C.B.C. and F.F.S. performed the measurements. I.C.B.C., F.F.S. and J.F. analysed and interpreted the data. I.C.B.C., F.F.S. and J.F. derived the theory. All authors discussed the results and contributed to the writing of the manuscript.

\textbf{Competing interests} The authors declare no competing interests.

\textbf{Reprints and permissions} To obtain permission to re-use content from this article visit RightsLink.

\textbf{Corresponding author} Correspondence to Ileana-Cristina Benea-Chelmus (ileanab@ethz.ch) or J\'er\^ome Faist (jerome.faist@phys.ethz.ch).

\section*{Methods}
\textbf{Signal demodulation at $\frac{f_{rep}}{2}$.} The measurement of the electro-optic correlation function on vacuum fields presented in this work required measurement integration times that exceeded $10^4$~s per measurement point. To ensure that the sole noise present in the system was shot noise, we implemented an effective noise cancelling technique. It consisted of the demodulation of the balanced voltages from the two balanced photodetectors at half the repetition rate of the femtosecond laser oscillator prior to the computation of the correlation function. A phase locked demodulation is algebraically equivalent to the subtraction of  measurements from temporally adjacent femtosecond probing pulses. This technique preserves the measurement of the electro-optic field correlation function, provided that the sampled THz fields are coherent only on time scales shorter than $T_{rep}=1/f_{rep}=12.5~ns$, which is fulfilled in the present case:
\begin{align}
\begin{split}
&\langle(\hat S_{eo}(t, \vec r)-\hat S_{eo}(t+T_{rep}, \vec r)) (\hat S_{eo}(t+\tau, \vec r + \delta \vec r)-\hat S_{eo}(t+\tau+T_{rep}, \vec r + \delta \vec r)) \rangle= \\
&\langle \hat S_{eo}(t, \vec r)\hat S_{eo}(t+\tau, \vec r + \delta \vec r) \rangle + \langle \hat S_{eo}(t+T_{rep}, \vec r)\hat S_{eo}(t+T_{rep}+\tau, \vec r + \delta \vec r) \rangle .
\end{split}
\label{eq:RFref}
\end{align}

The cross-terms  $\langle \hat S_{eo}(t, \vec r)\hat S_{eo}(t+\tau+T_{rep}, \vec r + \delta \vec r)\rangle = \langle \hat S_{eo}(t+T_{rep}, \vec r)\hat S_{eo}(t+\tau, \vec r + \delta \vec r)\rangle= 0$. 

This technique presents two advantages: it removes any coherent nonlinear signals and suppresses the 1/f noise, in particular slow drift contributions. The noise properties of the algorithm are found in sec.~3.2 and 3.3 of the supplementary information.

\textbf{Calculation of photon number for 45~K data.} The number of photons per mode can be extracted from the Fourier transformation of the electro-optic correlation function. For this, we make use of equation (2). Since the crystal is cooled to 4~K, and only the radiation is at 45~K, the responsivity function $R(\Omega)$ is equal for the two measurements at 4~K and 45~K. Denoting $S_{G^{(1)}_{eo,4K}}(\Omega)$ the power spectrum at 4~K and $S_{G^{(1)}_{eo,45K}}(\Omega)$ the power spectrum at 45~K, we find  

\begin{equation}
\langle \hat n (\Omega) \rangle = \frac{1}{2}(\frac{ S_{G^{(1)}_{eo,45K}}(\Omega)}{ S_{G^{(1)}_{eo,4K}}(\Omega)}-1)
\end{equation}

\begin{figure*}[hbtp]
  \centering 
  \DIFaddendFL \includegraphics[width=8.9cm]{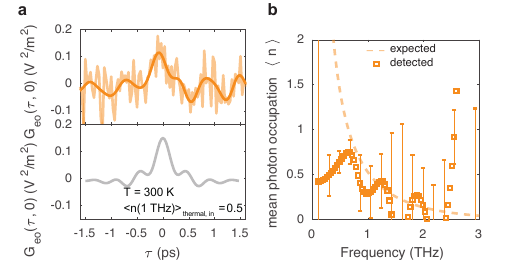}
  \caption{\textcolor{black}{\textbf{Electro-optic field correlation result of thermal radiation at 45~K.} \textbf{a,} Electro-optic field correlation measurement (top plot) is compared to the simulation (bottom plot). \textbf{b,} The mean number $\langle n \rangle$ of detected photons is compared to the expected mean number of photons at 45~K. For the estimation of the detected number of photons, see methods. In the region below 2~THz, only few photons per mode are detected. The error bars represent the  $1\sigma$ confidence interval.}  }
  \label{fig:45K}
\end{figure*}

To compute the photon number from Fig~\ref{fig:45K}b, we utilised the measured data at 45~K and 4~K.

\textbf{Data availability} The raw data associated with figures 2b, 2c, 3a, 3b, 3c, 3d, 3e, 4a, 4b are provided with the manuscript. Other data that support the findings of this study are available from the corresponding author on reasonable request.

\end{document}